\documentclass[12pt]{article}
\usepackage{amssymb,amsmath}
\usepackage[colorlinks=false,
      linkcolor=darkblue,  
      urlcolor=blue,    
      filecolor=blue,     
      citecolor=red,
linktocpage=true,
      pdfstartview=FitV,
      bookmarksopen=true]{hyperref}
\usepackage{latexsym}
\usepackage{graphicx}
\usepackage{subfigure}

\numberwithin{equation}{section}

\def\coeff#1#2{\relax{\textstyle {#1 \over #2}}\displaystyle}

\def\IR{\mathbb{R}}
\def\RR{\mathbb{R}}
\def\ZZ{\mathbb{Z}}
\def\Neql#1{{\cal N}\!=\!{#1}}

\def\cA{{\cal A}}

\def\cF{{\cal F}}

\def\cO{{\cal O}}

\newcommand{\dd}{\mathrm{d}}				

\newcommand{\F[1]}{F^{(#1)}}
\newcommand{\dF[1]}{\dot{F}^{(#1)}}
\newcommand{\vF[1]}{\vec{F}^{(#1)}}
\newcommand{\dvF[1]}{\dot{\vec{F}}^{(#1)}}

\newcommand{\omg}{\omega}

\newcommand{\bq}{\begin{equation}}
\newcommand{\eq}{\end{equation}}

\begin{document}  

\begin{titlepage}
 
\bigskip
\bigskip
\bigskip
\bigskip
\begin{center} 
{\Large \bf   Multi-Superthreads and Supersheets}

\bigskip
\bigskip 

{\bf 
Benjamin E. Niehoff, Orestis Vasilakis and Nicholas P. Warner  \\ }
\bigskip
\bigskip
Department of Physics and Astronomy \\
University of Southern California \\
Los Angeles, CA 90089, USA  \\
\bigskip
bniehoff@usc.edu,~vasilaki@usc.edu, ~warner@usc.edu  \\
\end{center}

\begin{abstract}

\noindent We obtain new BPS solutions of six-dimensional,  $\Neql 1$ supergravity coupled to a tensor multiplet.  These solutions are sourced by multiple ``superthreads" carrying D1-D5-P charges and two magnetic dipole charges. These new solutions are sourced by multiple threads with independent and  arbitrary shapes and  include new shape-shape interaction terms.   Because the individual superthreads can be given independent profiles,  the new solutions   can  be smeared together into continuous ``supersheets,'' described by arbitrary functions of two variables.  The supersheet solutions  have singularities  like those of the three-charge, two dipole-charge generalized supertube in five dimensions and we show how  such five-dimensional solutions emerge from a very simple choice of profiles.  The  new solutions obtained here also represent an important step in finding  superstrata,  which are expected to play a role in the description of black-hole microstates, due to their ability to store a large amount of entropy in their two-dimensional profile. 
\end{abstract}

\end{titlepage}


\tableofcontents

\section{Introduction}

The last six months have led to some very interesting new developments for BPS solutions in six dimensions.  First, the BPS equations for six-dimensional, minimal $\Neql 1$ supergravity \cite{Gutowski:2003rg} coupled to an anti-self-dual tensor multiplet \cite{Cariglia:2004kk} were shown to be linear\footnote{As with the corresponding result in five dimensions \cite{Bena:2004de, Bena:2007kg}, the equations that determine the spatial base geometry are still non-linear.}  \cite{Bena:2011dd}.  This not only provides a huge simplification in  solving these equations but it also enables one to use superposition to obtain multi-component solutions and, more abstractly, analyze the moduli spaces of such solutions.  It is not only anticipated that this will lead to interesting new developments in the study of black-hole microstate geometries but that it will also lead to interesting new results for holography on $AdS_3 \times S_3$ geometries.

Another interesting development is the conjecture that, in six dimensions, there is a new class of  BPS microstate geometries that depend upon several functions of two variables \cite{Bena:2011uw}.  These {\it superstrata}  generalize supertubes in several important ways.   Supertubes are supergravity backgrounds that carry two electric charges and one magnetic dipole charge and, in the IIB duality frame,   are completely smooth solutions  \cite{Lunin:2001jy,Lunin:2002iz}.   It is such geometries that lie at the heart of Mathur's original fuzzball proposal for the microstate structure of two-charge black holes (see, for example, \cite{Mathur:2005zp,Chowdhury:2010ct}). While there has been a great deal of progress on microstate geometries for the three-charge system (see \cite{Bena:2007kg, Mathur:2008nj, Balasubramanian:2008da, Skenderis:2008qn} for some reviews),  superstrata could lead to microstate geometries that provide the dominant semi-classical contribution to the microstate structure of the three-charge system.  The conjectured superstratum carries {\it three}  electric charges and three dipole charges, two of which are independent, and is described by an arbitrary, $(2+1)$-dimensional world-surface in six space-time dimensions. 

The argument for the existence of the superstratum has its origins in earlier work \cite{deBoer:2010ud} that suggested that one should be able to make two independent supertube transitions to produce new BPS solutions that carry three electric charges, two magnetic dipole charges and depend upon functions of two variables.  It was originally believed that such objects would be non-geometric and have spatial co-dimension two, but it was shown in \cite{Bena:2011uw} that  if  one does this in the proper manner for the D1-D5-P system in IIB supergravity then the result will  not only be a {\it geometric} BPS object with co-dimension {\it three} but one that is also completely smooth. Indeed, very near the superstratum the geometry approaches  that of the  supertube and so the  smoothness follows  directly from that of the supertube geometry.  Thus the superstratum provides a new microstate {\it geometry} of co-dimension three that carries three electric charges, two independent magnetic dipole charge and depends upon several functions of two variables

While the arguments given in  \cite{Bena:2011uw} for the existence of the superstratum are fairly compelling, it still remains to construct one explicitly and thereby establish its existence beyond all doubt.   The fact that the BPS equations in six dimensions are  linear gives one hope that the explicit supergravity solution may just be within reach (although it will still be extremely complicated).   The construction in  \cite{Bena:2011uw} has the virtue that it  lays out a sequence of steps,  via two supertube transitions, to arrive at the superstratum and so a possible  route to making a superstratum might be to replicate these intermediate steps  in a series of progressively more complicated but exact supergravity solutions.  Indeed some initial  progress in this direction was achieved in  \cite{Bena:2011dd} where the  D1-D5-P  system was pushed through the first supertube transition to obtain a new  three-charge, two-dipole charge\footnote{The two dipole charges in this solution are related to one another and so, to get to the superstratum, a further independent dipole charge must be added via a second supertube transition  \cite{Bena:2011uw}.}  generalized supertube  with an arbitrary profile as a function of one variable. We  will refer to such a solution  as a {\it superthread}. 

The next step towards a superstratum, which will be the subject of this paper, requires the construction of a multi-superthread solution  that could then be smeared  to a continuum and thus obtain a three-charge solution with a two-dimensional spatial profile that is a function of two variables.    This solution will still be singular and, like the standard supertube, will only become regular after the second supertube transition in which a Kaluza-Klein monopole is combined with the smearing.  This last step is probably going to be  the most difficult and will not be addressed here.  

\begin{figure}[!ht]
\begin{center}
\includegraphics[width=8.0cm]{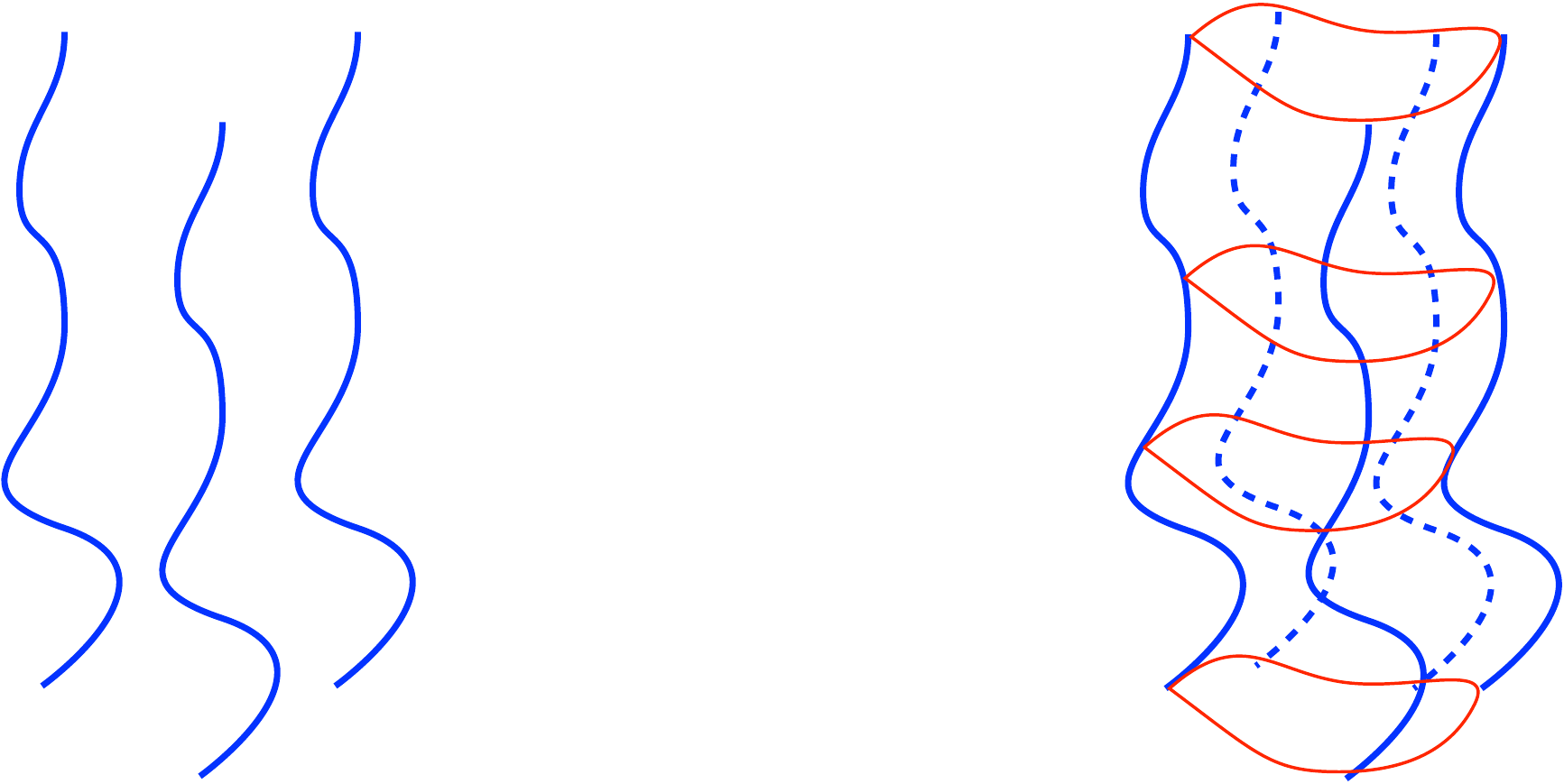}
\caption{ \small \it  Multi-thread solution in which all the threads are parallel.  When smeared the sheet profile is described by a product of functions of one variable:  the original thread profile and the thread densities.}
\label{ParallelThreads}
\end{center}
\end{figure}

\begin{figure}[!ht]
\begin{center}
\includegraphics[width=8.0cm]{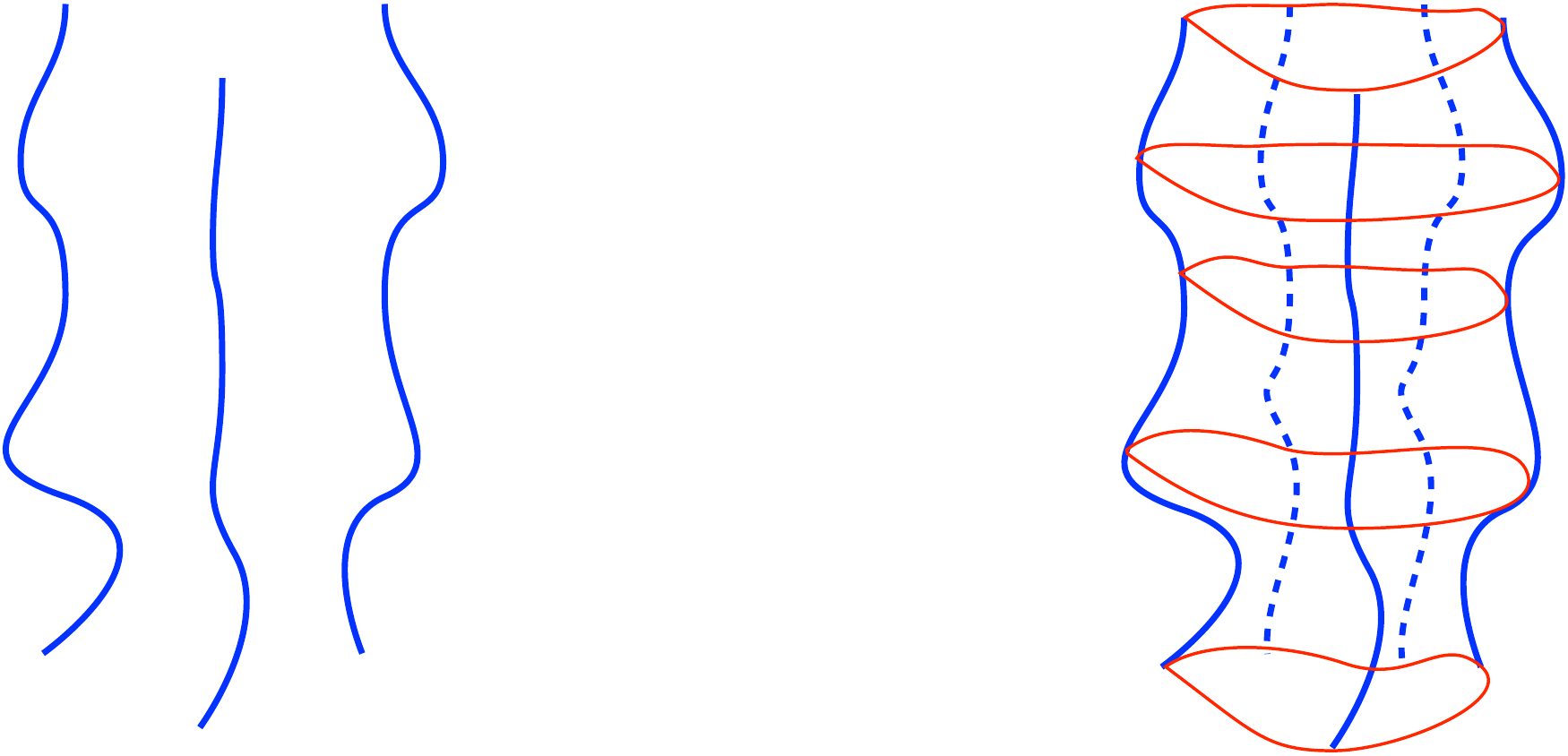}
\caption{ \small \it     Multi-thread solution in which all the threads have independent profiles.  When smeared the sheet profile is described by generic functions of two variables.}
\label{IndepThreads}
\end{center}
\end{figure}

In   \cite{Bena:2011dd} the step to the multi-superthread was only achieved for the highly restricted situation in which each thread was given exactly the same profile with a rigid translation to each distinct center. The smearing of such a multi-threaded solution will thus produce two-dimensional surface that is determined by a several functions of \emph{one} variable, namely  the smearing density and the original superthread profile functions. (See Fig. \ref{ParallelThreads}).  To get a surface that is truly a generic  function of two variables one must find the multi-superthread solution in which the threads at each center have  independent profile  functions so that, in the continuum limit, one obtains a one-parameter family of curves and hence  a surface swept out by a generic function of two variables  (See Fig. \ref{IndepThreads}).  The purpose of this paper is to find this general  a multi--thread solution.  The difficulty that we overcome here is that multiple superthreads with different profiles have highly non-trivial shape-shape interactions and we show exactly how these contribute to the angular momentum and local momentum charge densities.  It should, of course, be stressed that even though our solutions represent only a step towards the ultimate goal of the superstratum, the multi-superthread solutions presented here are completely new BPS solutions that are interesting in their own right.

In Section 2 we will briefly summarize the linear BPS system that we must solve and Section 3 we briefly summarize the superthread of  \cite{Bena:2011uw} and then present the new family of solutions.

\section{Solving the BPS equations}

\subsection{The BPS equations}

The six-dimensional metric  has the form:
\begin{equation}
ds^2 =  2 H^{-1} (dv+\beta) \big(du + \omega ~+~ \coeff{1}{2}\, \cF\, (dv+\beta)\big) ~-~  H \, ds_4^2\,.   \label{sixmet}
\end{equation}
where the  metric on the four-dimensional base  satisfies some special conditions that will not be relevant here because we are going to take $ds_4^2$ to be the flat metric on $\IR^4$. The coordinate, $v$, has period $L$ and the metric functions generically depend upon both $v$ and the $\IR^4$  coordinates, $\vec x$. We are also going to take the trivial fibration by setting $\beta =0$, which means, in particular, that there will be no Kaluza-Klein monopoles in the solution.  The spatial part of the metric  is simply flat $\IR^4 \times S^1$.

With these choices, the first step in constructing a solution is to determine the harmonic functions that encode the D1 and D5 charges:
\begin{equation} 
 *_4 \tilde d *_4 \tilde d \, Z_a ~=~  -\nabla^2 Z_a  ~=~  0 \,, \qquad i  =1,2   \,, \label{eoms1}
\end{equation} 
where $\tilde d$ is the exterior derivative and $\nabla^2$ is the Laplacian on $\IR^4$.  The metric function, $H$, is then given by  $H=\sqrt{Z_1Z_2}$ and one then must find self-dual Maxwell fields
\begin{equation} 
 \Theta_a  ~=~ *_4  \Theta_a\,, \qquad i  =1,2        \,, \label{sdfields}
\end{equation} 
that satisfy 
\begin{equation}
\tilde d  \Theta_1 ~=~   \coeff{1}{2} \, \partial_v \big[*_4 (\tilde d   Z_2) \big]  \,, \qquad \tilde d  \Theta_2 ~=~    \coeff{1}{2} \, \partial_v \big[  *_4 (\tilde d   Z_1) \big]      \,.  \label{eoms2}
\end{equation}
These Maxwell fields determine the magnetic components  of the fluxes in six dimensions and thus the magnetic dipole  D1 and D5 charges.

The angular momentum vector is obtained by solving
\begin{equation}
 (1+*_4)\, \tilde d\omega ~=~ 2\, (Z_1\Theta_1+Z_2\Theta_2) \, .\label{eoms3}
\end{equation}
and the last metric function is determined via: 
\begin{equation}
 *_4\tilde d *_4 \tilde d \,\cF ~=~   -\nabla^2 \cF  ~=~    2*_4\tilde d *_4 \dot \omega    -2 (\dot Z_1 \dot Z_2+Z_1\ddot Z_2+\ddot Z_1 Z_2)  +4*_4(\Theta_1\wedge \Theta_2)   \,. \label{eoms4}
\end{equation}
This last function encodes the momentum (P) charge of the solution.

\subsection{The new solutions}

The first steps in our new solution directly parallel those of  \cite{Bena:2011dd}.  
The harmonic functions, $Z_i$,   are  sourced on the thread profiles, $\vec F^{(p)}(v)$: 
\begin{align}
 Z_i=  1 ~+~ \sum_{p=1}^n{Q_{i\, p}\over  |\vec x-\vec F^{(p)}(v) |^2} \,,
\label{ZansatzD1D5P}
\end{align}
where we have required that $Z_i \to 1$ at infinity so that the metric is asymptotically Minkowskian.  The Maxwell fields, $\Theta_i$, that solve  (\ref{sdfields}) are simply given by:
\begin{equation}
 \Theta_i  ~=~ \coeff{1}{2}\, (1+*_4)\,\tilde d \, \bigg(\sum_{p=1}^n{Q_{i\, p}\, \dot{F}^{(p)}_m dx^m \over |\vec x-\vec F^{(p)}(v) |^2}  \bigg ) \, .
 \label{Thetaex}
\end{equation}
As noted in  \cite{Bena:2011dd}, the magnetic dipoles of this solution may be thought of as being defined by
\begin{equation}
\vec d_1 ~=~  Q_1  \dot{\vec F} (v) \,, \qquad \vec d_2 ~=~  Q_2 \dot{\vec F} (v) \,,  \label{dipoles} 
\end{equation}
and they satisfy the constraint that is familiar from the five-dimensional, generalized supertube   \cite{Bena:2004wt,Bena:2004wv,Bena:2005ni}:
\begin{equation}
Q_1 \, \big | \vec d_2\big | ~=~  Q_2 \, \big | \vec d_1\big | \label{diprelns} \,.
\end{equation}
This means that even though the solution has two dipole charges, only one of them is independent of the other charges. 

To write the solution for the angular momentum vector and the third function, $\cF$, it is useful to define:
\begin{equation}
\vec R^{(p)}   ~\equiv~  \vec x ~-~ \vec F^{(p)}(v)  \,, \qquad R_p   ~\equiv~ \big | \vec R^{(p)}   \big |  ~\equiv~ \big| \vec x ~-~  \vec  F^{(p)}(v)    \big| \, ,
 \label{Rdefns}
\end{equation}
and for each $p$ and $q$, introduce the anti-self-dual two form area element:
\begin{align}
\cA^{(p,q)}_{ij}    ~\equiv~  R^{(p)}_i R^{(q)}_j   - R^{(p)}_j R^{(q)}_i  ~-~  \varepsilon^{ijk\ell} R^{(p)}_k R^{(q)}_\ell  \, ,
 \label{Apqdefns}
\end{align}
where $\varepsilon^{1234} = 1$.  The angular momentum vector can be written in three pieces:
\begin{equation}
\omega ~=~ \omega_0  ~+~ \omega_1   ~+~ \omega_2\, .
 \label{omegabits}
\end{equation}
where the first two parts are very similar to the those in \cite{Bena:2011dd}:
\begin{align}
 \omega_0  ~=~  &   \sum_{i=1}^2 \,\sum_{p=1}^n{Q_{i\, p}\, \dot{F}^{(p)}_m dx^m \over |\vec x-\vec F^{(p)}(v) |^2} \,,   \nonumber \\
 \omega_1  ~=~ &   \frac{1}{2}  \,\sum_{p, q=1}^n \, (Q_{1\, p}\,Q_{2\, q} +Q_{2\, p}\,Q_{1\, q})  \, { \dot{F}^{(p)}_m dx^m \over R_p^2  \,R_q^2} \,.  
 \label{omega01defn}
\end{align}
The last part of the solution, $\omega_2$, is  part of our new result and arises from the interaction between non-parallel threads:
\begin{equation} 
\omega_2 ~=~   \frac{1}{4}  \,\sum_{p, q=1 \atop p \ne q}^n  (Q_{1p} Q_{2q} + Q_{2p} Q_{1q}) \,  \frac{\big( \dot{F}^{(p)}_i - \dot{F}^{(q)}_i \big)  }{\big | \vec F^{(p)}  - \vec F^{(q)} \big | ^2} \, \bigg\{ \bigg( \frac{1}{R_p^2} - \frac{1}{R_q^2} \bigg)  \, dx^i
~-~ \frac{2}{R_p^2 R_q^2} \, \cA^{(p,q)}_{ij}  \, dx^j  \bigg\} \,.
 \label{omega2defn}
\end{equation}

From this one can easily verify that 
\begin{equation}
\vec \nabla \cdot \vec \omega ~=~- \partial_v (Z_1 Z_2) \,,
 \label{divom}
\end{equation}
which means that the equation for $\cF$ simplifies to
\begin{equation}
\begin{split}
\nabla^2 \cF & ~=~  - 2\, \big [\dot Z_1 \dot Z_2~+~  *_4(\Theta_1\wedge \Theta_2) ~\big]   \\
& ~=~  - 4\,  \sum_{p, q=1}^n  (Q_{1p} Q_{2q} + Q_{2p} Q_{1q})  \, \frac{1}{R_p^4 R_q^4} \, \Big[\,  (\vec R^{(p)}  \cdot \vec R^{(q)} )\,  \Big( {\dot {\vec {F}}}{}^{(p)} \cdot {\dot {\vec {F}}}{}^{(q)}  \Big)  ~-~ {{\dot {\vec {F}}}{}^{(p)}}{}^i {{\dot {\vec {F}}}{}^{(q)}}{}^j \, \cA^{(p,q)}_{ij}  \, \Big]     \,.
 \end{split}
  \label{Feqn1}
\end{equation}
This can be solved by the somewhat obvious guess:
\begin{equation}
\begin{split}
\cF  ~=~  -4~-~ 4  \sum_{p=1}^n \,  \frac{Q_{3\, p}}{R_p^2}   ~-~&  \frac{1}{2} \, \sum_{p, q=1  }^n    \, \frac{(Q_{1p} Q_{2q} + Q_{2p} Q_{1q})}{R_p^2 R_q^2} \, \Big( {\dot {\vec {F}}}{}^{(p)} \cdot {\dot {\vec {F}}}{}^{(q)}   \Big)   \\
 + ~ & \sum_{p, q=1 \atop p \ne q }^n  (Q_{1p} Q_{2q} + Q_{2p} Q_{1q})  \, \frac{1}{R_p^2 R_q^2} \,  \frac{  \dot{F}^{(p)}_i \, \dot{F}^{(q)}_j \,  \cA^{(p,q)}_{ij} }{\big | \vec F^{(p)}  - \vec F^{(q)} \big | ^2}   \,,
 \end{split}
  \label{Fsol1}
\end{equation}
where the first two terms represent particular choices for the harmonic pieces of $\cF$.    In normalizing these harmonic pieces we have   kept in mind the fact that  dimensional reduction to five space-time dimensions yields $\cF =   -4   Z_3$, where $Z_3$ determines the third electric charge of the solution and is on the same footing (in five dimensions) as $Z_1$ and $Z_2$.  The terms in $\omg$ and $\mathcal{F}$ that contain $\cA^{(p,q)}_{ij}$ express the non-trivial interaction between non-parallel superthreads. These terms vanish for solutions with multiple threads of parallel profiles, $\vec{F}(v)$, and hence did not appear in \cite{Bena:2011dd}.

Finally, there are also possible harmonic pieces that can be added to the angular momentum vector, $\omega$.  To define these, introduce the following self-dual harmonic forms on $\IR^4$:
\begin{equation}
\begin{split}
\Omega^{(1)}_+  & ~=~ dx^1 \wedge dx^2 ~+~   dx^3 \wedge dx^4  \,, \\
 \Omega^{(2)}_+   & ~=~ dx^1 \wedge dx^3 ~-~   dx^2 \wedge dx^4  \,, \\ 
\Omega^{(1)}_+  & ~=~ dx^1 \wedge dx^4 ~+~   dx^2 \wedge dx^3  \,.
 \end{split}
  \label{sdforms}
\end{equation}
Then the following are zero modes of the equation (\ref{eoms3}) that defines $\omega$:
\begin{equation}
 \omega_{harm}  ~=~    \sum_{a=1}^3\,\sum_{p=1}^n \, {1\over R_p^4}\, J_{ p}^{(a)}(v)  \,  \Omega^{(a)}_{+ ij} \,  R^{(p)  i}  dx^j  \,,
  \label{omharm}
\end{equation}
where the $J_{ p}^{(a)}(v)$ are $v$-dependent angular momentum densities.  The one-form in (\ref{omharm}) is sourced along the profile of the superthread.  Moreover, one can easily verify that:
\begin{equation}
*_4 \tilde d *_4  \omega_{harm}  ~=~    0 \,,
  \label{omharmLor}
\end{equation}
and so this induces no additional contribution to $\cal F$ in  (\ref{eoms4}).
 
\subsection{Regularity and the near-thread limit}
The six-dimensional metric  we are considering is:
\begin{equation}
ds^2 ~=~ 2 (Z_1   Z_2)^{-{1 \over 2}}  dv \,  \big(du + \omega ~+~ \coeff{1}{2}\, \cF\, dv \big) ~-~  2 (Z_1 Z_2)^{{1 \over 2}} \, |d \vec x |^2  \,.   \label{simpsixmet}
\end{equation}
Regularity requires that $Z_1 Z_2 >0$ and we will ensure this by taking
\begin{equation}
Q_{1 p}\, , \, Q_{2  p}  ~\geq~ 0  \,,  \label{poschgs}
\end{equation}
for all $p$.  Moreover, if one sets all displacements to zero except along the circular fiber parametrized by $v$ then the metric collapses to $ds^2 = (Z_1   Z_2)^{-{1 \over 2}}  \cF\, dv^2$, which means that one must require 
\begin{equation}
- \cF  ~\geq~ 0  \label{negF}
\end{equation}
everywhere if one is to avoid closed timelike curves.  The expression for $\cF$ in   (\ref{Fsol1}) is somewhat complicated but the condition (\ref{negF}) can generically be satisfied if one takes $Q_{3p}$ to be positive and large enough.  We will discuss this further below. 

The near-thread limit is going to be singular because it is locally a three-charge, two-dipole charge object.   However we must also ensure that there are no closed time-like curves (CTC's) near the superthreads.  To that end we collect all the divergent and finite parts of the metric in the limit $R_p \to 0$:
\begin{equation}
\begin{split}
Z_i  & ~\sim~  {Q_{i p} \over R_p^2}  ~+~  Q_{i p}  ~+~ \sum_{q \ne p}  \,{Q_{i q} \over R_{pq}^2}  ~+~ \cO(R_p) \,, \qquad  i = 1,2  \,, \\
{\cal F}  & ~\sim~  -  {  Q_{1 p}  \, Q_{2 p}\over R_p^4} \, \big| {\dot {\vec {F}}}{}^{(p)} \big|^2  ~-~{1  \over R_p^2}  \bigg[ 4\, Q_{3 p}  ~+~  \sum_{q \ne p}  \,{(Q_{1 p} Q_{2 q} + Q_{2 p} Q_{1 q})\over F_{pq}^2}   \,  \big( {\dot {\vec {F}}}{}^{(p)} \cdot {\dot {\vec {F}}}{}^{(q)}\big)   \bigg] ~+~ \cO(1) \,, \\ 
\omega  & ~\sim~  -  {  Q_{1 p}  \, Q_{2 p}\over R_p^4} \,\big( {\dot {\vec {F}}}{}^{(p)} \cdot d \vec x \, \big)   ~+~  {1\over R_p^3}\,  \sum_{a=1}^3\,  J_{ p}^{(a)}(v)  \,  \Omega^{(a)}_{+ ij} \,   \widehat R^{(p) i}  \,    dx^j  \\
&  \qquad ~+~  {1\over R_p^2} \,  \bigg[ (Q_{1 p} + Q_{2 p})   ~+~ \sum_{q \ne p}  \,{(Q_{1 p} Q_{2 q} + Q_{2 p} Q_{1 q})\over F_{pq}^2}     \bigg]  \, \big( {\dot {\vec {F}}}{}^{(p)} \cdot d \vec x \, \big)     ~+~  \cO \Big( {1\over R_p}\Big)     \,.
 \end{split}
  \label{nearasymp}
\end{equation}
where we have included the harmonic pieces, (\ref{omharm}),  of $\omega$ and where
\begin{equation}
 F_{pq}^2   ~\equiv~\big | \vec F^{(p)}  - \vec F^{(q)} \big | ^2  \,, \qquad \widehat R^{(p)} ~\equiv~ { \vec R^{(p)}  \over R_p}  \,.
\label{defns1}
\end{equation}

Setting $du =0$, one finds, at  leading order as $R_p \to 0$,
\begin{equation}
ds^2 ~\sim~  {\sqrt{Q_{1 p} Q_{2 p} } \over R_p^2}\, \Bigg[ \big| {\dot {\vec {F}}}{}^{(p)} \big|^2  \Bigg(\, dv - { {\dot {\vec {F}}}{}^{(p)} \cdot d \vec x  \over \big| {\dot {\vec {F}}}{}^{(p)} \big|^2}   \Bigg)^2 ~+~ dx_\perp^2 \, \Bigg]    \,.   \label{metasymp1}
\end{equation}
where
\begin{equation}
dx_\perp^2  ~\equiv~ |d\vec x|^2  ~-~ {  \big|{\dot {\vec {F}}}{}^{(p)} \cdot d \vec x \big|^2  \over \big| {\dot {\vec {F}}}{}^{(p)} \big|^2} \,,
\label{perpmet}
\end{equation}
which is the spatial metric in $\IR^4$ perpendicular to the tangent, ${\dot {\vec {F}}}{}^{(p)}$, to the superthread.  The asymptotic metric (\ref{metasymp1}) is manifestly positive but not positive-definite:  There is a null direction along the supertube.  That is, the leading order terms vanish precisely if one takes 
\begin{equation}
d \vec x ~=~  {\dot {\vec {F}}}{}^{(p)}  \, d \lambda  \,, \qquad  d v ~=~ \big| {\dot {\vec {F}}}{}^{(p)} \big|  \, d \lambda \,,
\label{nullrir}
\end{equation}
for some infinitesimal displacement, $d \lambda$.

For this displacement one finds a leading order term coming from the harmonic pieces  of $\omega$:
\begin{equation}
ds^2_\lambda~=~     {d \lambda^2     \over R_p}\, {1\over\sqrt{Q_{1 p} Q_{2 p} }} \,  \sum_{a=1}^3\,  J_{ p}^{(a)}(v)  \,  \Omega^{(a)}_{+ ij} \,   \widehat R^{(p) i}  \,   {\dot { {F}}}{}^{(p) j} 
\label{subleading1}
\end{equation}
If one looks in the direction $R^{(p)}_i  \sim - \sum_{a=1}^3\,  J_{ p}^{(a)}(v)  \,  \Omega^{(a)}_{+ ij} \,  {\dot { {F}}}{}^{(p) j} $ one finds that $ds^2_\lambda$ is negative and proportional to $\sum_{a=1}^3\,  (J_{ p}^{(a)}(v) )^2  \big| {\dot {\vec {F}}}{}^{(p)} \big|^2$.  Thus for a superthread with ${\dot {\vec {F}}}{}^{(p)} \ne 0$ one can only avoid CTC's if one sets 
\begin{equation}
J_{ p}^{(a)}(v)   ~=~   0  \, \label{noharm}
\end{equation}
that is, the harmonic pieces, (\ref{omharm}), produce CTC's and so must be discarded.  The complete physical solution is thus given by  $\omega_0 + \omega_1 + \omega_2$ defined in (\ref{omega01defn}) and  (\ref{omega2defn}).

An important consequence of this analysis  is that the angular momentum vector is completely determined by the electric charges and profiles of the configuration.  This is slightly different from the five-dimensional solutions in which one has independent choices of harmonic functions in the angular momentum vectors and the angular momenta are then fixed in terms of the charges and positions of the sources via bubble equations, or integrability conditions, that remove CTC's.  For the six-dimensional solutions presented here one fixes charges, positions and profiles and the angular-momentum vector is adjusted automatically:  there are no bubble equations.

Having now killed the leading order of the metric along the displacement (\ref{nullrir}) it turns out that there is a finite order piece.  As $R_p \to 0$ the metric becomes:
\begin{equation} \label{subleading2}
\begin{split}
ds^2_\lambda~=~     {d \lambda^2     \over \sqrt{Q_{1 p} Q_{2 p} }} \,  \bigg[&  4\, Q_{3 p}   - \big|\dvF[p]\big|^2 ( Q_{1p} + Q_{2p} ) \\  
& ~-~  \sum_{q \ne p}  \,{(Q_{1 p} Q_{2 q} + Q_{2 p} Q_{1 q})\over F_{pq}^2}   \,  {\dot {\vec {F}}}{}^{(p)} \cdot \big(  {\dot {\vec {F}}}{}^{(p)}  -  {\dot {\vec {F}}}{}^{(q)}\big) \bigg] ~+~ \cO(R_p) .
\end{split}
\end{equation}
Again, to avoid closed timelike curves we require that the quantity in brackets be non-negative, which is equivalent to asking that
\begin{equation} \label{subleading3}
- \cF ~\geq~  \dF[p]_i \, \omega_i
\end{equation}
near each thread.  Hence the positivity of $ds_\lambda^2$ in (\ref{subleading2}) places a lower bound on each of the charges $Q_{3p}$: 
\begin{equation} \label{bound1}
 Q_{3 p}  ~\geq~  \frac{1}{4}\,   \big|\dvF[p]\big|^2 ( Q_{1p} + Q_{2p} )  ~+~ \frac{1}{4}\, \sum_{q \ne p}  \,{(Q_{1 p} Q_{2 q} + Q_{2 p} Q_{1 q})\over F_{pq}^2}   \,  {\dot {\vec {F}}}{}^{(p)} \cdot \big(  {\dot {\vec {F}}}{}^{(p)}  -  {\dot {\vec {F}}}{}^{(q)}\big)  \,.
\end{equation}
The individual bounds  for each $p$ depend upon the detailed geometric layout of the threads  but if one sums  over all the threads then one obtains a global bound upon the total charges:
\begin{equation} \label{bound2}
 \sum_{p=1}^n \, Q_{3 p}  ~\geq~  \frac{1}{4}\,    \sum_{p=1}^n \,  \big|\dvF[p]\big|^2 ( Q_{1p} + Q_{2p} )   \,.
\end{equation}

The origins of these bounds can be understood in terms of ``charges dissolved in flux'' \cite{Bena:2004de}.  From (\ref{dipoles}) one sees that the right-hand sides of  (\ref{bound1}) and  (\ref{bound2}) can be thought of as the dipole-dipole interactions that give rise to an effective  electric contribution to the Kaluza-Klein charge described by $\cF$.   As we will describe below, the harmonic charge term, described by $Q_{3p}$ in $\cF$, is the charge measured at infinity and so these bounds mean that the only physically sensible solutions are those in which one does indeed correctly account, at infinity, for the charge coming dipole-dipole interactions.

\subsection{Asymptotic charges}

The electric charges measured at infinity come from the asymptotic forms of $Z_1$, $Z_2$ and  $Z_3 \equiv -\frac{1}{4}  \cF$.  From  the leading ($\cO(R^{-2})$) terms in  (\ref{ZansatzD1D5P})  and (\ref{Fsol1}) one can easily read off the  D1, D5, and P charges: 
\begin{equation} \label{D1D5P charges}
\text{D1:}   \ \  \sum_p Q_{1p}, \qquad \text{D5:} \ \ \sum_p Q_{2p}\, \qquad \text{P:} \ \  \sum_p Q_{3p}.
\end{equation}
The  terms in the tensor, $\cA^{(p,q)}_{ij}$,  defined  (\ref{Apqdefns})  do not contribute in $\cF$ because $\vec R^{(p)}$ and $\vec R^{(q)}$  become nearly parallel at large distances and so this term vanishes at leading order.

The asymptotic form of $\omega$ can be massaged into
\begin{equation} \label{asymomega}
\begin{split}
\omega &\sim \frac{1}{R^2} \sum_p \big( Q_{1p} + Q_{2p} \big) \, \dvF[p] \cdot \dd \vec x + \frac{2}{R^4} \sum_p \big( Q_{1p} + Q_{2p} \big) \big( \vec R \cdot \vF[p] \big) \, \dvF[p] \cdot \dd \vec x \\
& \qquad \qquad + \frac12 \frac{1}{R^4} \sum_{\substack{p, q \\ p \neq q}} \frac{Q_{1p} Q_{2q} + Q_{2p} Q_{1q}}{F_{pq}^2} R^i \bigg[ \F[pq]_i \, \big( \dvF[pq] \cdot \dd \vec x \big) \\
& \qquad \qquad \qquad \qquad \qquad - \dF[pq]_i \, \big( \vF[pq] \cdot \dd \vec x \big) + \varepsilon_{ijk\ell} \F[pq]_j \dF[pq]_k \, \dd x^\ell \bigg] \\
& \qquad \qquad + \frac12 \frac{1}{R^4} \vec R \cdot \dd \vec x \, \sum_{\substack{p, q \\ p \neq q}} \frac{Q_{1p} Q_{2q} + Q_{2p} Q_{1q}}{F_{pq}^2} \big( \vF[pq] \cdot \dvF[pq] \big),
\end{split}
\end{equation}
where $\vF[pq] \equiv \vF[p] - \vF[q]$.  The first term falls of as $R^{-1}$ and is perhaps somewhat unexpected.   Mathematically it arises through the contribution of the constant terms in the $Z_i$ to the source for $\omega$ in (\ref{eoms3}).  These source terms mean that, to leading order, $ (1+*_4)\tilde d \omega$ limits to $2(\Theta_1+ \Theta_2)$ and thus $\omega$ inherits an asymptotic behavior given by the vector fields in parentheses in (\ref{Thetaex}).  In five dimensions, $(\Theta_1+ \Theta_2)$ falls off  faster and leads to standard expansions for angular momenta in $\omega$.  The presence of the $\cO(R^{-1})$ terms in six-dimensions comes because of the $v$-dependent sources in  (\ref{eoms2}).  The fact that this term is a total $v$-derivative means it will always vanish when we reduce to five dimensions.  This is because, in order to reduce to five dimensions, the sources must be smeared in a way that kills all $v$ dependence; hence the unusual $\cO(R^{-1})$ term disappears and one recovers the standard behavior of five-dimensional solutions.  We will illustrate this in the next section.

Physically, the  $\cO(R^{-1})$ terms represent  a \emph{linear} momentum for the configuration.  The somewhat unusual  feature of the six-dimensional linear system is that all the equations are solved on a constant-$v$ slice and that, for a given value of $v$, the solution is insensitive to the configuration at other values of $v$ and so, slice-by-slice, the solution sees the superthread as indistinguishable from the   thread that carries a linear momentum.  It is only when one smears the solution along a closed profile that the solution combines different sections of the solution with different orientations so that the leading momentum behavior cancels and leaves one with a more standard angular momentum.

The second term is in (\ref{asymomega}) is purely rotational, and expresses the difference $J_T \equiv J_1 - J_2$.   The third term is the potential of a purely anti-self-dual 2-form, and so it expresses the sum $J_1 + J_2$.    The last term is a total derivative, and may be viewed as pure gauge.   

\section{Supersheets}
\label{Sect:Sheets}

\subsection{General supersheets}

It is straightforward to take the continuum limit of the multi-superthread solution.  The set of profiles, $\vec F^{(p)}(v)$, are replaced by a function of two variables, $\vec F (\sigma, v)$,  the discrete charges,  $Q_{i \, p} $,  are replaced by density functions, $\rho_i(\sigma)$ and the sums are replaced by integrals. Thus we have
\begin{equation}
 Z_i=  1 ~+~ \int_0^{2\,\pi} \, {\rho_i(\sigma) \, d \sigma \over  |\vec x-\vec F (\sigma, v) |^2} \,, 
\label{contZi}
\end{equation}
\begin{equation}
 \Theta_i  ~=~ \coeff{1}{2}\, (1+*_4)\,\tilde d \, \bigg( \int_0^{2\,\pi} \, { \rho_i(\sigma) \,   \partial_ v  \vec  F(\sigma,v) \cdot  d\vec x   \over |\vec x-\vec F (\sigma, v)  |^2} \, d \sigma \bigg ) \,,
 \label{contTheta}
\end{equation}
where we have chosen to normalize the smearing over the interval $[0, 2\pi]$.
Following (\ref{Rdefns}) and  (\ref{Apqdefns}) we define 
\begin{equation}
\vec R (\sigma)   ~\equiv~  \vec x ~-~ \vec F (\sigma, v)   \,, \qquad R (\sigma)     ~\equiv~ \big |  \vec R (\sigma, v, \vec x) \big | \, ,
 \label{contRdefns}
\end{equation}
and the tensor
\begin{equation}
\cA_{ij}  (\sigma_1, \sigma_2)    ~\equiv~  R_i (\sigma_1)   R_j (\sigma_2)  -  R_j (\sigma_1)   R_i  (\sigma_2) 
   ~-~  \varepsilon^{ijk\ell} R_k (\sigma_1)  R_\ell   (\sigma_2)\, ,
 \label{contAdefns}
\end{equation}
With these definitions, the rest of the continuum solution can be written
\begin{align}
 \omega_0  ~=~  &   \sum_{i=1}^2 \, \int_0^{2\,\pi} \, {\rho_i(\sigma) \, \partial_v  \vec  F(\sigma,v) \cdot  d\vec x  \over |\vec x-\vec F (\sigma, v)  |^2}\, d \sigma \,,   \nonumber \\
 \omega_1  ~=~ &   \frac{1}{2}  \,\int_0^{2\,\pi}  \int_0^{2\,\pi} \,   (\rho_1(\sigma_1) \rho_2(\sigma_2)+\rho_2(\sigma_1) \rho_1(\sigma_2))  \, {\partial_v  \vec  F(\sigma_1,v) \cdot  d\vec x \over R (\sigma_1, v, \vec x)^2  \, R (\sigma_2, v, \vec x)^2} \, d \sigma_1 d \sigma_2  \,,  
 \label{contomega01}
\end{align}
\begin{equation} \label{contomega2}
\begin{split} 
\omega_2 ~=~     \frac{1}{4}  \,\int_0^{2\,\pi}  \int_0^{2\,\pi} \,   & (\rho_1(\sigma_1) \rho_2(\sigma_2)+\rho_2(\sigma_1) \rho_1(\sigma_2))  \,   \frac{\big( \partial_v  F_i (\sigma_1,v)  -   \partial_v  F_i (\sigma_2,v) \big)  } {\big |  \vec  F (\sigma_1,v)   - \vec F (\sigma_2,v) \big | ^2}  \\
&  \bigg\{ \bigg( \frac{1}{R (\sigma_1)^2} - \frac{1}{R (\sigma_2)^2} \bigg)  \, dx^i ~-~ \frac{2}{R (\sigma_1)^2  R (\sigma_2)^2 } \, \cA_{ij}  (\sigma_1, \sigma_2)  \, dx^j  \bigg\} \, d \sigma_1 d \sigma_2  \,,
\end{split}
\end{equation}
\begin{equation} \label{contFsol1}
\begin{split}
\cF  ~=~  &- 4~-~ 4\,\int_0^{2\,\pi}    \frac{ \rho_3(\sigma) }{R (\sigma)^2}  \, d \sigma \\
& -~ \int_0^{2\,\pi}  \int_0^{2\,\pi} \,  (\rho_1(\sigma_1) \rho_2(\sigma_2)+\rho_2(\sigma_1) \rho_1(\sigma_2))  \   \frac{1}{R (\sigma_1)^2  R (\sigma_2)^2 }  \,\\
& \qquad\qquad \Bigg[\,  \frac{1}{2} \, \big( \partial_v  \vec F  (\sigma_1,v)\big)   \cdot \big(\partial_v  \vec F  (\sigma_2,v)  \big) ~-~    \frac{  \partial_v F_i(\sigma_1,v) \,  \partial_v F_j (\sigma_2,v) \ \,  \cA_{ij}  (\sigma_1, \sigma_2) }{\big |  \vec  F (\sigma_1,v)   - \vec F (\sigma_2,v) \big |^2}  \, \Bigg]  \, d \sigma_1 d \sigma_2  \,. 
\end{split}
\end{equation}

The integrals for $\omega_2$ and $\cF $ have potential singularities at the coincidence limits, $\sigma_1 = \sigma_2$, with a double pole coming from the denominator factor of $ |  \vec  F (\sigma_1,v)   - \vec F (\sigma_2,v)  |^{2}$.  However, the tensor $\cA_{ij}$ has a simple zero as $\sigma_1 \to \sigma_2$ and  this skew tensor is further contracted with factors that have simple zeroes in the coincidence limit.  Thus there is also  a double zero in the numerator  leading to a finite contribution in the coincidence limit.

While we have smeared the multi-superthread solution into a single supersheet, it is also clear that one can smear the multi-superthread solutions into multiple supersheets and such solutions will be given by straightforward generalizations of (\ref{contZi})--(\ref{contFsol1}).

Finally, we note that one can, of course, recover the multi-superthread solutions from this continuum solution by replacing the density functions, $\rho_a$, by sums over delta functions:
\begin{equation}
\rho_a(\sigma)  ~=~  \sum_{j =1}^N \, Q_{a\, p} \, \delta \big(\sigma - \sigma^{(p)} \big) \,, \qquad a=1,2,3  \,,
 \label{rhodelta}
\end{equation}
and where the individual profile functions are specified by the sampled values of $\vec F (\sigma,v)$:
\begin{equation}
\vec {F}{}^{(p)}(v)  ~=~  \vec F \big ( \sigma^{(p)},v\big)  \,.
 \label{Fps}
\end{equation}
%

\subsection{The five-dimensional generalized supertube as a supersheet}

The supersheets described above are sourced by sheet profiles described by arbitrary functions of \emph{two} variables and are thus much more general than previously-known solutions.  However, it is worthwhile to smear our solutions in a more trivial way in order to see exactly how  five-dimensional solutions emerge.  Therefore we give an example that produces a $v$-independent sheet profile, allowing us to reduce on the $v$ fiber and obtain a standard five-dimensional solution.

A useful, non-trivial  way to accomplish this is to choose any profile $\vec{F}(\sigma)$ in $\RR^4$, and define $\vec F   ( \sigma,v ) =  \vec F  ( \sigma+ \kappa v) $.  The result  should then be a solution of the standard, linear BPS system in five dimensions  \cite{Bena:2004de, Bena:2007kg}.  One should also directly recover physical constraints like radius relations.  
For  simplicity, we will take the charge densities to be constant and we will smear a simple helical configuration that will produce a cylinder along $v$ and a ring in $\IR^2 \subset \IR^4$:
\begin{equation}
\vec F ( \sigma,v )~=~  \Big(0,0,a \cos (\kappa v + \sigma), a \sin (\kappa v + \sigma) \Big) \,,
 \label{simpF}
\end{equation}
where $\kappa$ and $a$ are constants with $\kappa= \frac{2 n\pi}{L}$, for $n\in \ZZ$.  Each thread will have a constant charge distribution, given by
\begin{equation}
\rho_i (\sigma) \equiv \frac{Q_i}{2 \pi}.
\end{equation}

To carry out the integrals \eqref{contZi}, \eqref{contomega01}, \eqref{contomega2}, \eqref{contFsol1}, it is easiest to work in polar coordinates on $\RR^2 \times \RR^2$ given by:
\begin{equation}
x^1 = \eta \cos \phi, \qquad x^2 = \eta \sin \phi, \qquad x^3 = \zeta \cos \psi, \qquad x^4 = \zeta \sin \psi. \label{polar basis}
\end{equation}
Note that $R^2 = \eta^2 + \zeta^2$.
From these coordinates we can easily go to spherical coordinates by defining $\eta= R \cos \theta$ and $\zeta =R \sin\theta$.  Then, for example, we obtain $Z_1$ by integrating
\begin{equation}
\begin{split}
Z_1 &= 1 + \frac{Q_1}{2 \pi} \int_0^{2 \pi} d\sigma \, \frac{1}{\eta^2 + \zeta^2 + a^2 - 2 a \zeta \cos (\sigma + \kappa v)} \\
&= 1 + \frac{Q_1}{\sqrt{(\eta^2 + \zeta^2 + a^2)^2 - 4 a^2 \zeta^2}}.
\end{split}
\end{equation}
The rest of the integrals are tedious, but straightforward.  The result is
\begin{gather} 
Z_{1,2} = 1 + \frac{Q_{1,2}}{\Sigma}, \qquad \cF =  -4 - \frac{4 \, Q_3}{\Sigma} - \kappa^2 \, Q_1 Q_2 \, \frac{1}{\Sigma} \bigg( \frac{\eta^2 + \zeta^2}{\Sigma} - 1 \bigg), \\
\omega = \frac{\kappa}{2} \, (Q_1 + Q_2) \bigg( \frac{ \eta^2 + \zeta^2 + a^2}{\Sigma} - 1 \bigg) \, d \psi + \kappa \, Q_1 Q_2 \, \frac{1}{\Sigma^2} \, \big( \eta^2 \, d \phi + \zeta^2 \, d \psi \big),
\end{gather}
where we have defined
\begin{equation}
\Sigma \equiv \sqrt{(\eta^2 + \zeta^2 + a^2)^2 - 4 a^2 \zeta^2} ~=~ \sqrt{(R^2 + a^2)^2 - 4 a^2 R^2 \sin^2 \theta}.
\end{equation}

At infinity, these behave as
\begin{align} 
\label{ringasymp}
Z_{1,2} & ~\sim~ 1 + \frac{Q_{1,2}}{R^2}\, , \qquad \qquad Z_3   ~=~  - \coeff{1}{4}\, \cF  ~\sim~ 1 +  \frac{ Q_{3}}{R^2}  \\
\omega & ~\sim~ \frac{\kappa }{R^2}  \big[\big ( (Q_1 + Q_2)  \, a^2 +  Q_1 Q_2  \big) \sin^2 \theta\, d \psi   +   Q_1 Q_2 \,   \cos^2 \theta \, d \phi  \big] \\
& ~=~ \frac{1}{2\, R^2} \big[J_1 \, \sin^2 \theta\, d \psi   ~+~   J_2 \,  \cos^2 \theta \, d \phi  \big]   \label{angmoms1} \,,
\end{align} 
where the five-dimensional angular momentum vector, $k$, is related to $\omega$ via $\omega = 2 k$.  This explains the factor of $2$ in (\ref{angmoms1}).

This solution corresponds, as expected,  to the three-charge, two-dipole-charge generalized supertube \cite{Bena:2005ni}, with charges $Q_1, Q_2, {Q}_3$, and dipole charges
\begin{equation} \label{prop}
q^1 \equiv - \frac{\kappa Q_2}{2}, \qquad q^2 \equiv - \frac{\kappa Q_1}{2}, \qquad q^3 \equiv 0.
\end{equation}
We define $\widetilde{Q}_3$ as
\bq \label{tildeQ3}
\widetilde{Q}_3 ~\equiv~ Q_3-\coeff{1}{4} \, \kappa^2  Q_1Q_2
\eq
Note that $\widetilde  Q_3$ is  the constituent electric charge while the charge measured at infinity, $Q_3$, also contains the charge arising from the dipole-dipole interaction.

From (\ref{angmoms1}) one can read off the angular momenta and  one can also check that the radius relation:
\begin{equation}
J_T  ~\equiv~  J_1 -J_2 ~=~ -\coeff{1}{2} \, \kappa\, a^2 \,  (Q_1 + Q_2) ~=~  (q^1 + q^2 + q^3) \, a^2
\end{equation}
is  satisfied automatically.

The condition that one has $\cF \le 0$ globally implies that $\widetilde{Q}_3 \ge 0$ and hence: 
\begin{equation}
\label{Fnegsmeared}
 Q_3~ \ge~\coeff{1}{4} \,  \kappa^2\,Q_1Q_2  ~=~ q^1 \, q^2
\end{equation}
This is simply the continuum analog of (\ref{bound2}).

Near the ring, we find that to avoid CTC's  one must have:
\begin{equation} \label{smearedCTC1}
Q_1Q_2\, \big(\widetilde{Q}_3 -\coeff{1}{4} \,  \kappa^2a^2(Q_1+Q_2)\big)  ~\ge~ 0 \,, 
\end{equation}
and hence
\begin{equation} \label{smearedCTC2}
 \widetilde{Q}_3 ~\ge~  \coeff{1}{4} \,  \kappa^2a^2 \,(Q_1+Q_2)   ~=~ \coeff{1}{2} \,  \kappa  \, J_T    \,.
\end{equation}

This is not quite the same as the continuum limit of (\ref{bound1}) because the latter bound was derived assuming that $R_q$ remained finite as $R_p \to 0$ whereas the continuum limit gets other important terms in from the coincidence limits when two threads approach one another.  This is evident from the fact that the general integrals in Section \ref{Sect:Sheets} are finite in the coincidence limit  but the continuum limit of (\ref{bound1})  involves a divergent integral.

We have thus recovered one of the standard five-dimensional solutions.  The process of obtaining a solution in five dimensions usually involves choosing some harmonic functions  and then adjusting the coefficients so as to avoid closed timelike curves.  These choices are already implicit in our six-dimensional solution and emerge directly in the smeared solution.

\section{Conclusions}

The BPS equations in six-dimensional, minimal $\Neql 1$ supergravity coupled to one tensor multiplet have been shown to be a linear system \cite{Bena:2011dd} once an appropriate base geometry has been determined.  This allows one to use superposition to create a wide variety of solutions and such solutions could lead to interesting new developments in the study of black hole microstate geometries, as well has holography on $AdS_3 \times S_3$. It has also been conjectured \cite{Bena:2011uw} that a new class of BPS microstate geometries,  \emph{superstrata},  may exist.  Such objects carry three electric charges and two \emph{independent} dipole charges, depend on arbitrary functions of two variables and  are expected to be \emph{regular} solutions in the IIB duality frame.   They are thus a sheetlike, three-charge generalization of the supertube.   The fact that they depend upon functions of two variables suggests that they should be able to store large amounts of entropy in their shape modes, indeed the superstrata microstate geometries are expected to give the dominant semi-classical contribution to the entropy of the three-charge system.

While compelling arguments have been given for the existence of superstrata \cite{Bena:2011uw}, it remains to explicitly construct one.  The results we present here are a very significant step in that  direction. 

The non-trivial aspect of our new solutions  is that they take into account the shape-shape interactions of the separate superthreads. It was evident in \cite{Bena:2011dd} that  superthreads  interact non-trivially with one another when the threads have different profiles and so the completely general multi-superthread was not constructed. Indeed, as depicted in Fig. \ref{ParallelThreads}, the multi-centered solutions found  in \cite{Bena:2011dd}  only involves parallel  threads,  shifted by rigid translation in $\RR^4$.  Such solutions can only be smeared together into a sheet depending on arbitrary functions of \emph{one} variable with one set of functions describing the thread profile and another defining the smearing densities.  To get a solution that is genuinely a function of two variables by smearing, it is essential to construct the multi-superthread solution in which all the threads can have independent profiles and so the smeared threads yields a thread-density profile, $\vec F(\sigma, v)$. This is depicted in  Fig. \ref{IndepThreads}.

In this paper we have analyzed the effect of this shape-shape interaction and presented the general solution with multiple threads of completely arbitrary and independent shapes at each center.  These  solutions were then smeared to obtain new solutions sourced by a two-dimensional sheet of completely arbitrary profile, described by arbitrary functions of \emph{two} variables.  It is also evident that our results can easily be generalized to multi-supersheet solutions.

We also checked our results against a known five-dimensional solution by taking  a simple helical  profile and smearing it to a cylindrical sheet and dimensionally reducing.  We thus recovered the generalized supertube solution with  three-charges and two-dipole charges  \cite{Bena:2004wt,Bena:2004wv,Bena:2005ni}.  We found that   CTC conditions, like the radius relation, which  usually require an additional constraint on the five dimensional solution,  emerge automatically from our six-dimensional solutions.  

The solutions presented in this paper are completely new geometries and are interesting in their own right as three-charge solutions sourced by arbitrary two-dimensional surfaces.  To obtain the superstratum we will need to do exactly what we have achieved here but with an additional KKM magnetic charge smeared along the profile thereby providing the required second \emph{independent} dipole moment \cite{Bena:2011uw}.  We leave this for future work.

\section*{Acknowledgements}

We would like to thank Iosif Bena for very valuable conversations. This work is supported in part by DOE grant DE-FG03-84ER-40168. OV would like to thank the USC Dana and David Dornsife College of Letters, Arts and Sciences for support through the College Doctoral Fellowship and the USC Graduate School for support through the Myronis Fellowship.




\end{document}